\documentclass[journal,12pt,onecolumn,draftclsnofoot,]{IEEEtran}
\usepackage{cite}
\usepackage{amsmath,amssymb,amsfonts}
\usepackage{algorithmic}
\usepackage{graphicx}
\usepackage{textcomp}
\usepackage{xcolor}
\usepackage{gensymb}
\usepackage{ragged2e}
\usepackage{comment}

\justifying
\renewenvironment{IEEEbiography}[1]
  {\IEEEbiographynophoto{#1}}
  {\endIEEEbiographynophoto}

\begin{document}

\title{On the Mutuality between Localization and Channel Modeling in sub-THz}

\author{Eray~Güven,~\IEEEmembership{Student Member,~IEEE,}
        Güneş~Karabulut-Kurt,~\IEEEmembership{Senior Member,~IEEE}

 \thanks{E. Güven and G. Karabulut-Kurt are with the {Department of Electrical Engineering,  Polytechnique Montr\'eal, Montr\'eal, Canada, e-mails: \{guven.eray, gunes.kurt\}@polymtl.ca}.}}


\maketitle

\begin{abstract}
The use of sub-Terahertz (sub-THz) band is gaining considerable attention in 6G networks. In this study, we introduce hardware and propagation integrated \textit{3D Propagation Model} to describe sub-THz channels and discuss its advantages over both deterministic and stochastic 6G channel models. The unexplored mutuality of localization and communication is presented and its potential in \textcolor{black}{integrated sensing and communication} (ISAC) applications \textcolor{black}{is} highlighted. Afterward, a real-time sub-THz localization experiment \textcolor{black}{is} conducted to show the impact of the mispositioned and misaligned narrow beams on service quality. In continuation, we highlight the most current challenges and developments in THz localization \textcolor{black}{and explore the potential of sub-THz frequencies to efficiently utilize the ultra-wideband spectrum}. In the end, the open issues that need to be overcome to provide high spatial resolution and millidegree-level angle of arrival estimation in ISAC applications have been explored. 

\end{abstract}

\section{Introduction}

The sub-Terahertz (sub-THz) spectrum between 90 GHz and 300 GHz is a driving force in the transition to very high data rate networks as well as a bridge to \textit{True THz} region which has Tbps level data rate. 5G New Radio applications for the intelligent mobility management models are not sufficient enough to answer high rate, secure and robust for sustainable sensing and localization systems which leads to dead-ends and dilemmas. Such an example, it \textcolor{black}{is} shown that ultra reliable and low latency (URLLC) brings device compatibility and leads to performance limits, enhance mobile broadband (eMBB) applications does not meet the real time protocols for Open Systems Interconnection (OSI), or massive machine type communication (mMTC) is constrained by the power limitations for large scale deployments. On top of these, one technology limits the other and the use of incoming THz bands are limited by the THz gap as there is no \textcolor{black}{alternative solution} for the time being. 

The sub-THz operating band \textcolor{black}{holds the potential to propel 6G} networks out of this rabbit hole in a small time. One prominent advantage of sub-THz over the middle and high end of the THz band is that \textcolor{black}{the sub-THz band operates at a low enough frequency to avoid significant exposure to molecular absorptions} \cite{han2022molecular} yet, high enough to adopt flat phased Gaussian beam wavefront to provide efficient propagation, mode matching and high degree of coherence. Thereupon, several reasons to utilize THz communication (THzCom) features are preserved by the sub-THz band as well such as having a better spatial resolution, reduced diffractions and enhanced penetration through materials. Over an extended period, THzCom became a breakthrough and a promising step for \textcolor{black}{integrated sensing and communication} (ISAC) applications. The most recent THz roadmap study \cite{leitenstorfer20232023} shows that even though there are some proof-of-concept examples that can operate in 0.1-10 THz with above 10 Gbps data rate, signal-to-noise ratio (SNR) and phase noise performances are still critical points that need to be overcome. Sub-THz not only enables a very high data rate ($\sim$ 1 Tbps) within the International Radio Regulation (RR) defined range between 90 GHz and 275 GHz but also provides an escape way for unreachable hardware related THz band challenges. To set an example for an indoor localization opportunity, sub-THz radiation exhibits moderate absorption characteristics in many common materials, enabling its propagation through objects that would otherwise be opaque to visible light or conventional radio waves where the data rate is very limited. Furthermore, it allows field measurement techniques \cite{kanhere2021outdoor} for positioning and ranging on top of the conventional time of flight (ToF) and phased-based ranging techniques. On a material level, it \textcolor{black}{is} shown that sub-THz field power is stable and linear for nanocomposites and polymers, unlike the True THz region where the field power drops to $10^{-12}$ level with poor consistency. 

\textcolor{black}{Channel modeling, mathematically describing the wave propagation, is critically important for the determination of signal quality, capacity and communication risks. Sub-THz channel modeling differs from microwave modeling due to the increased path loss, distinct antenna structure, and heightened susceptibility to factors such as molecular absorption and scattering. Therefore, channel characteristics such as delay spread, channel gain, and coherence time are strongly dependent on hardware capabilities and the propagating environment. This study puts a novel perspective for channel modeling in THzCom localization.} \textcolor{black}{A few number of propagation paths, the impact of antennas on deterministic channel models becomes significant for high accuracy. From the localization perspective, the contribution of the very directive antennas contributes the most. A channel model taking the antennas into account for localization application not only bounds the localization capability of the systems but also can be used primarily as a localization method. As an example, phase shifting on an antenna simply can change the fading model due to the very high angular resolution of the THz antennas. The sensing capabilities of THz propagation can be exposed in localization with only channel state information (CSI).} \textcolor{black}{Figure \ref{alld}(a) illustrates a scenario where a transmitter (BS) localizes users with only CSI.}

The contributions of this study are the following:
\begin{itemize}
    \item A novel hardware integrated THz channel modeling approach has been introduced. A truly realistic and non-complex channel definition is provided by including THz antennas in the channel model, moving away from deterministic or stochastic scaling. 

    \item The mutual relationship between localization and communication \textcolor{black}{is} constructed and the possibilities that this inclusive mutuality would bring are explored. 

    \item A real-time sub-THz experiment \textcolor{black}{is} conducted to illustrate the importance of localization and the complexity of mispositioning. Observations in 3 different sub-THz frequencies are evaluated with in-phase and quadrature (IQ) based performance analysis.

    \item Open issues related to artificial intelligence (AI), ISAC and hardware in the THz band have been raised. Methods to solve the double-edged problems are discussed. 

\end{itemize}

\section{3D Propagation Model}
\label{3dpm}

\begin{figure*}[t] 
\centering
    \includegraphics[width=1\textwidth]{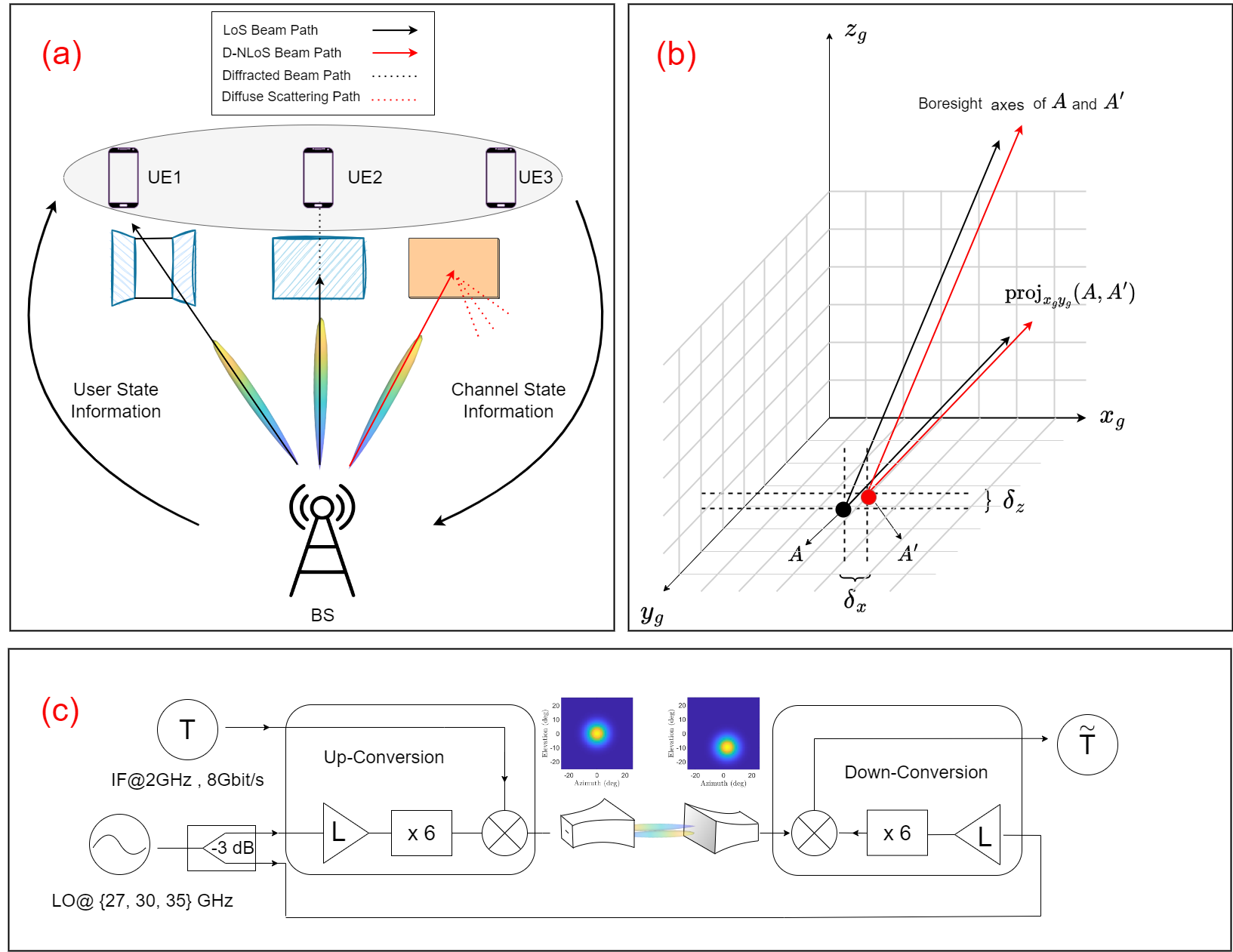}
    \caption{A sub-THz localization application portrait. \textcolor{black}{ In (a), the unique feedback of exposing directive antennas in localization for LoS - near line-of-sight - NLoS scenarios respectively. Thus, even a non-existent USI gives a position information. In (b), a mispositioned antenna boresight axes are depicted as implemented in the experiments. In (c), the multi-sub-THz band experimental setup diagram with hardware and beam configuration is shown.}}
    \label{alld}
\end{figure*}

Statistical physical channel modeling for the radio frequency is a quite common and generalizable way to describe wave propagation behavior under various environments. \textcolor{black}{For example, a deterministic fluctuating two-ray (FTR) channel model is a proper solution for not only pencil beam waves but also lower band mm-Wave}. However, an imperfect directional antenna involved in the channel model along with misalignment, mispositioning and non-linearity in wave sources brings a new aspect and challenges to the THz channel modeling. Sub-THz and higher bands are exposed to $<1$ ms coherence time and notably higher Doppler spread than mm-wave even if the BS and receiver (UE) are stationary. Accordingly, these networks require high rate synchronization, equalization, resource management and handover techniques. Despite this, antenna response time (ART) for on-board sensors is a significant problem at the front end. This delay is not part of the ToF, yet it is included in the propagation. If the process is linear, ART can be solved by calibration; otherwise, it's an error source that permanently exists. 

The illustration of mispositioning without misalignment in the transmitter \textcolor{black}{is} given in Figure \ref{alld}(b). A mobile UE with the targeted path node $A$ position in the global frame and mispositioned in $A'$ instead, without any additional error. Note that the positional offsets with $\delta_{x,z}$ correspond to a different set of phase and magnitude in the non-uniform field distribution of the receiver. Thence, the beam structure disforms in the global coordinate frame even though the transmitter-receiver beam shape is permanently fixed in their local coordinate frame. \textcolor{black}{Figure \ref{alld}(c) shows that sub-THz transmission with a precisely aligned matching antenna result in degrading the SNR because of the $10$ and $15$ degree rotational misalignment at the receiver in $y$ and $z$ axes respectively. The same figure illustrates the beam intensity in the transmitter and receiver body frame.} Mobile networks suffer the most due to this deterioration. The ineliminability of this error necessitates the inclusion of both BS and UE aperture radiation in the channel model. Briefly, antenna characteristics that are negligible for sub-6Ghz and mm-wave cannot be neglected for narrow beam THz.\textcolor{black}{ In the mitigation of the localization error for a single input single output (SISO), beam manipulation without an external tool is not feasible. Therefore, error mitigation can be autonomously achieved through mechanical means such as adaptive control of users and transmitters. A strategic approach is to establish an error budget, allowing the communication system a certain margin of error, and then implementing resource optimization within this budget. For instance, as the error decreases, power allocation can be reduced. 
In multiple input multiple output (MIMO), several options are viable. For example, a robust mitigation to localization error is the beamforming for SNR maximization.}

Although one recent study highlights the use of contemporary geometry based stochastic spatio-temporal channel model for 6G networks \cite{chen2019time}, we introduce a new paradigm of 3D Propagation Model (3DPM) that will include the radiator components into the channel description to provide a more generic joint channel and propagation model for THz networks. A 3DPM gives the probabilistic definition of an end-to-end sub-THz channel state under hardware characteristics such as polarization loss, antenna defectivity, asymmetric grating lobes, boresight misalignment, non-synchronized local oscillator (LO) sources, amplifier non-linearity and illumination efficiency if there is.

Considering the loss figures caused by the phase error of a standard gain antenna, defections on each plane lower the radiation resistance multiplicatively. The importance of the topic of interest lies in its deterministic process. Any misalignment and mispositioning will perturb the targeted antenna gain. As a result, the term ``radiated power" loses its reason, as it will result in a non-probabilistic deceptive gain. 3DPM eliminates the labor of such ill-advised solutions that will solve this issue, such as pre-calibration.

Forming the 3DPM is quite straightforward and it is no more than custom tailored finite-path channel derivation. In the case of AFP on the large scale fading process, gain components of the path loss model are simply extended with the AFP of each plane. In small scale fading, each line-of-sight (LoS) and scattered ray is associated with elevation ($\theta$) and azimuth ($\phi$) within UE local frame, governed by AFP($\theta,\phi$).

\section{THz Gap and Distributed Localization}
A pleasing development is that the THz Gap is gradually becoming out of interest topic for THz generators \cite{leitenstorfer20232023}. The current developments in linear THz generators and low-noise THz detectors cover the high phase noise and low power weaknesses each year. In below 300 GHz, silicon CMOS could have reached only $\sim 0.8$ mW, however with the current Indium Phosphide based techniques can go up to 1.17 mW/$\mu$m \cite{arabhavi2022inp}, while Galium-Phosphide frequency multipliers can achieve 200 mW \cite{siles2018new} under 500 GHz with larger bandwidth range. The state-of-the-art Silicon Germanium (SiGe) limitations of nearly 300 GHz in transition frequency show itself, while semiconductor lasers have the capability to wave generation starts with 30 THz \cite{zimmer2021sige}. In the experimental case, this gap gets wider, yet this gap gets shorter with recent pulse THz generators from 5 to 20 THz \cite{le2023novel}. 



As a consequence, a 0.3-30 THz gap in the electromagnetic (EM) spectrum is not a deadly challenge anymore, yet there is no practical way around to deal with sampling and signal processing such a high rate of data in wireless communication, leading to inevitable latency where it shatters the URLLC applications. \textcolor{black}{Nevertheless, solving the sampling issues introduces a breakthrough novelty called distributed localization (DL) \cite{roumeliotis2002distributed}.}

\textcolor{black}{The current centralized localization is heavily based on navigation satellites. In DL, each communication node localizes each node with respect to other nodes, effectively eliminating the need for a reference satellite and reducing the path loss exponentially. This development opens the door for the use of THzCom in DL. THzCom advantages are driven by two key factors: 1) ultra-wideband communication increases the localization accuracy and 2) THzCom enables the 3DPM mutuality for the localization. Thus, as long as THzCom is active, localization remains possible and vice versa. Provided that issues related to sampling rate and hardware are resolved, THzCom in DL represents a groundbreaking advancement where the ultra-fast sampling rate leads to an ultra-fast refresh rate for localization. The refresh rate for current GNSS-based localization is limited to 1 Hz.}


\section{Localization and Communication Conjunction}
\label{loc}

In an application specific perspective with a constant transmission window, the pointwise beam quality ($M$) factor depends on the detector position as the molecular absorption varies. Additionally, 3D AFP manipulates the transmission for directive antennas. The narrower the angular spread, the more sensitive the beam alignment. In short, the neglectable phased array focusing flaws becomes more critical. Whatsmore, these errors are propagated due to more difficult beam management tasks in wide beam applications. The asymmetrical side lobes with shifted boresight axis are a drastic reason that degrades the THz propagation. Note that this is not an exceptional situation for the THz region, but it also becomes crucial in both higher frequencies where atmospheric and molecular losses are severe and mobile applications. For example, in augmented and virtual reality localization, wave penetration is a cutting edge for sub-THz over light detection and ranging (LiDAR). LiDAR's inefficiency during heavy rains and under strong scintillation of sun angles can be compensated with sub-THz without losing accuracy and data rate.

\begin{figure}[] 
  \centering  \includegraphics[width=0.5\textwidth]{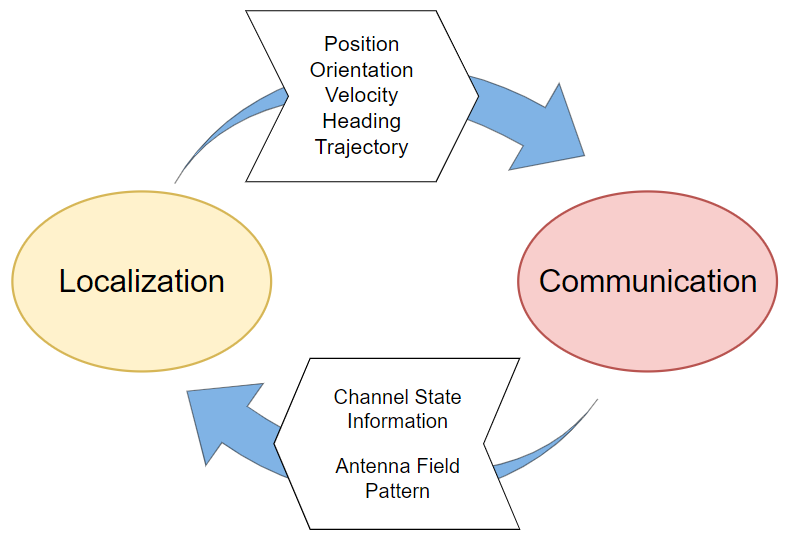}
    \caption{The localization and communication mutuality on THzCom. \textcolor{black}{THzCom and localization mutually necessitate and reinforce one another.}}
    \label{locztn}
\end{figure}

We highlight the dilemma and/or mutual relationship between localization and THzCom. Figure \ref{locztn} illustrates this relationship between the localization process and the sub-THz detector function. We define this symbiotic relation with mutuality in THz localization (MTL). Figure \ref{locztn} represents a functor $F:C\rightarrow D$ such that an object $F(x)$ in $D$ corresponds to every object $x$ in $C$. This opens a window of opportunities in radio localization, and some of those will be defined in next. 
First, statistical channel state information (sCSI) acquisition by the position is utterly possible. Note that, with the usage of MIMO systems, sCSI becomes highly critical when the channel fading becomes nearly flat. An explicit example \textcolor{black}{is} shown in Figure \ref{alld}(a) illustrating that fading distribution and ray estimation with only sight information (line-of-sight (LoS) / near line-of-sight / non line-of-sight (NLoS)) can be done and the other channel properties without loss of generality. In THz, sight terminology for THz propagation \textcolor{black}{is} redefined and novel descriptions exist in the literature such as \textcolor{black}{directed non line-of-sight (D-NLoS)} \textcolor{black}{is} defined \cite{serghiou2022terahertz} for this band to imply that scattering and specular reflections unable to cover a direct link as NLoS does, however; D-NLoS can be coordinated with reflector elements such as reflected intelligent surface (RIS) units. Especially in the case of MIMO, inter-path multiplexing is limited due to the lack of spatial degree of freedom, usage of spherical waves is a reasonable way to recover this channel sparsity. Regardless, there is an unorthodox way to exploit this nature of THz.

Each state of a UE is an object to the localization process and
we call this set of states user state information (USI). Despite limited knowledge about the radio environment, CSI contributes an amount of understanding of the USI. Under the wide-sense stationary uncorrelated channel assumption, a vague USI is obtained by having weak NLoS or strong LoS CIR information at the transmitter. Although this does not explicitly reveal the USI such position or speed of UE, it is used in resource management operations. Sub-6GHz bands cannot go further than this, however, THz bands can. A discrete example is that, in the case of the THz blockage effect, CSI enables to provision of direct USI. Additionally, due to the low spatial degree of freedom, the communication link approaches NLoS in case of mispositioning. Thus, CSI provides direct USI at the transmitter. Therefore, USI allows transmission flexibility to provide physical link health and security. On another note, the tracking rate of the narrow beam purely depends on the coherence time which is too low that the THz channel can be considered as time-invariant. \textcolor{black}{In mobile networks, this timeframe becomes even more limited} and sparse channel tracking can be done as with low pilot overhead \cite{liao2021terahertz}. 

Second, MTL can be exploited with dead reckoning on the self-navigation system as $F:C\rightarrow D$ and $F^*:D^*\rightarrow C^*$. An exciting technique that implicitly utilizes the MTL is channel charting (CC), facilitating radio mapping of the surroundings using learning-based methods. \cite{studer2018channel}. Not only does CC provide an unsupervised learning powered multi antenna localization, but it also can be integrated with MTL giving additional sensing information to be included in the kinematic model, enhancing the fusion capabilities. 

Lastly, the recent trajectory estimator methods can make use of the spatial geometry to approximate what sCSI is going to be for each mobile user. This allows superior interference management and capacity maximization of a network. On top of ongoing research, trajectory optimization with MTL is possible to maximize the path and fading loss to be able to localize and localize to be able to maximize USI.

\begin{figure}[]
    \centering   \includegraphics[width=1\textwidth]{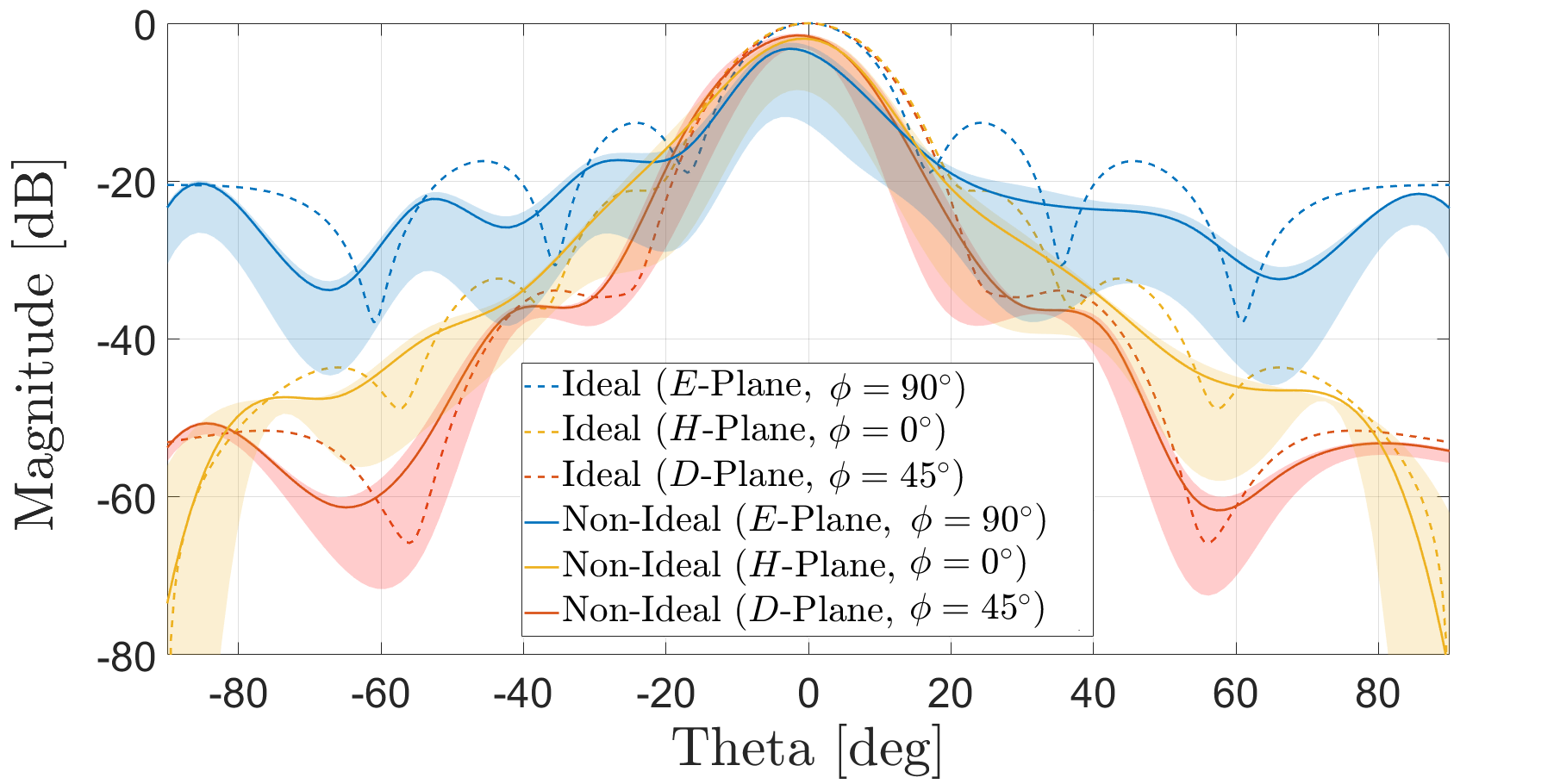}
    \caption{\textcolor{black}{A rectangular waveguide with 50 mils opening aperture (WR-5) nominal gain pattern and non-ideal WR-5 with \% 95 weighted confidence interval gain pattern in 182 GHz.}}
    \label{anten}
\end{figure}

\textcolor{black}{
A simple non-ideality on the hardware results in failed communication, unlike the rich scattering microwave communication. Whether utilizing deterministic or stochastic channel modeling, incorporating non-ideal hardware into the channel model establishes realistic accuracy bounds for THzCom. In localization, 3DPM offers unique benefits that aren't achievable with ideal hardware assumptions. On the downside, 3DPM necessitates hardware information, including the hardware non-idealities, which might be impractical for large-scale infrastructures.}

Considering the channel response in time varying fading, AFP also determines each multi-path components (MPC) contributions at different angle of arrival (AoA) and received power. Figure \ref{anten} illustrates an ideal circularly polarized G-band horn antenna field pattern (AFP) for each plane. Corresponding to this, a non-ideal AFP response with $\mathbb{N}(0,\sigma^2)$ impairment has been bounded with its weighted \%95 confidence interval. The 3D rectangular AFP of a standard gain G-band horn element antenna \textcolor{black}{is} shown in Figure \ref{anten}. While the antenna aperture efficiency (AE) is 0.511 as usual, the field distribution \textcolor{black}{becomes} distorted in higher frequencies and lowers the AE consequently. Hence, a saturation point exists to disprove the widely accepted assumption that the higher the frequency, the higher the gain. 

\section{Experimental Evaluation} 

In order to imply the significance of positioning in the sub-THz band, a short distance trial experiment \textcolor{black}{is} conducted. \textcolor{black}{Our demonstration contains 3 discrete operation points, namely $164$ GHz, $182$ GHz and $212$ GHz.} The SISO testbed contains one M8190A Analog Wave Generator (AWG), one E8267D Pulse signal generator (PSG) Vector Signal Generator, two VDI mm-Wave frequency extender with 140-220 GHz frequency range to block-up and block-down the Intermediate Frequency (IF) with a multiplier $N=6$ and lastly, one UXR0502A signal analyzer with 256 GSa/s capability. AWG is utilized to generate a modulated 2 GHz baseband signal and the PSG \textcolor{black}{generates} the 26-35 GHz LO source for both transceiver and detector. A wideband two-way splitter ($6-40$ GHz) \textcolor{black}{is} used to divide the source signal into VDIs. Thus, further synchronization problems are avoided and internal clock \textcolor{black}{is} used for wave generation. $1500$ frames for each observation \textcolor{black}{is deemed to be} sufficient enough to obtain average values. The identical aperture antennas with 8.9$^\circ$ and 10.28$^\circ$ of 3 dB beam width ($\theta_{3dB}$) on $E$-plane and $H$-plane respectively, follow the previously defined antenna model \textcolor{black}{is} used.

The cable and splitter sum loss on the testbed $\sim 11$ dB for each extender. The parameters of the different case scenarios are shown in Table 1. The focus of the experiment is to observe the position misalignment effects along $x^+$ and $x^-$-axes with $ \delta$ and $2 \delta$. This position error vector \textcolor{black}{is} implemented through 2-axis linear stage platforms carrying the VDI modules. Mean Error Vector Magnitude (EVM) measurements and EVM variance can be seen in Figure 5, along with the test parameters. Note that the elevation angles (pitch axis) of the modules are perfectly aligned. Neither modules nor antennas do not rotate. For simplicity, only the transmitter position is modified as a controlled variable.

\begin{figure}[] 
\centering
    \includegraphics[width=0.4\textwidth]{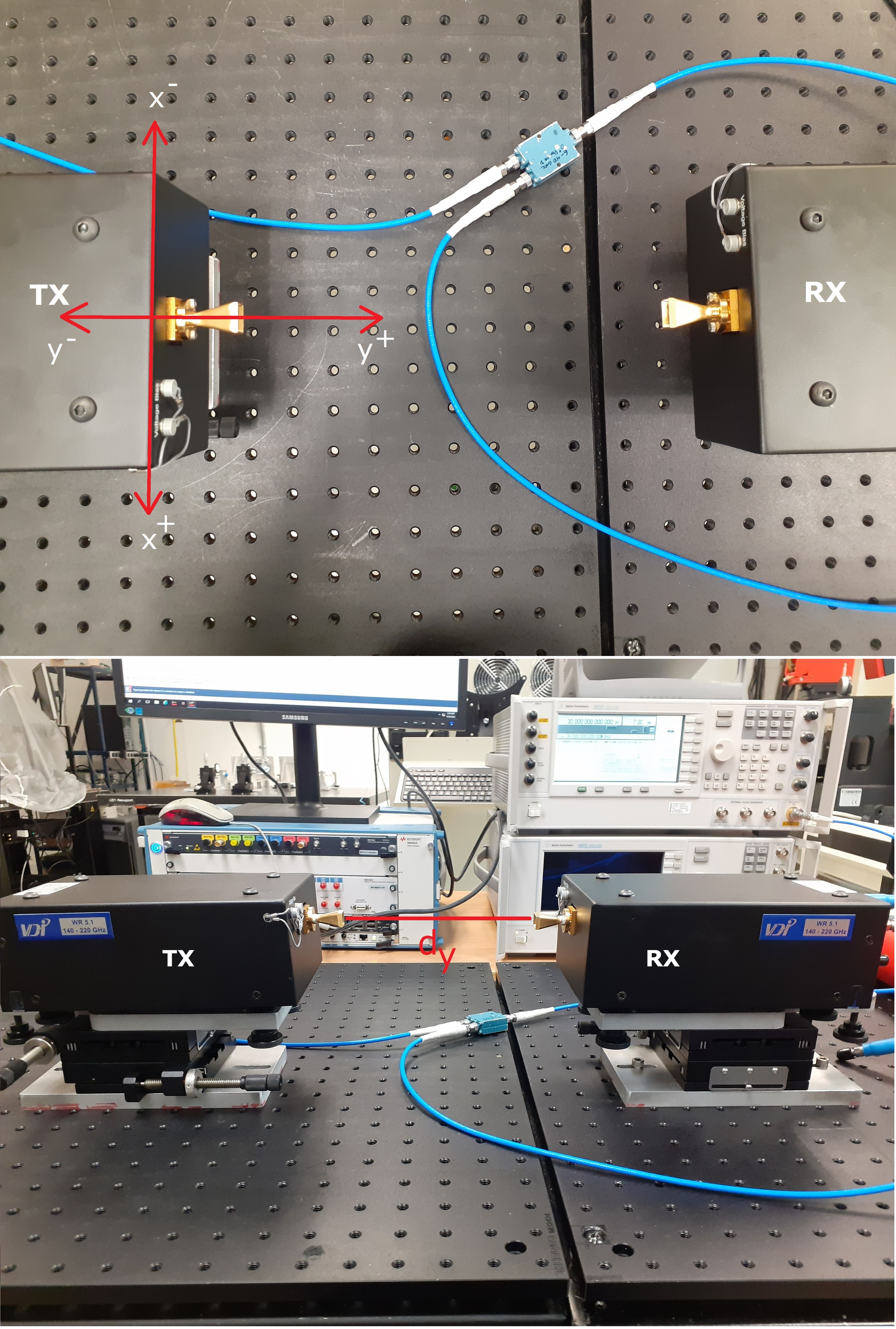}
    \caption{Ultra-wideband sub-THz testbed in PolyGrames.
The local coordinate space of Tx in the top photo illustrates the uniaxial motion. The $d_y$ = 22 cm indoor LoS link with 182 GHz is shown in the bottom photo. }
    \label{testtt}
\end{figure}

Circularly polarized $10$ dBi rectangular horn antennas are identical and cover the whole G-band. The same antennas have been used throughout all measurements. \textcolor{black}{The measurement setup satisfies the far-field distance $d_f < 0.177$ meter for each operating point.} The diagram of the testbed \textcolor{black}{is} depicted in Figure \ref{alld}(c) and the snapshots from the experiments can be seen in Figure \ref{testtt}.

It is observed that the transmission performance deteriorates as mispositioning accumulates in both axes. A notable observation is that mispositioning along $x^+$ and $x^-$ directions does not have the same deterioration for $\delta$ and $2\delta$ margins. One way of interpretation is that asymmetric side lobes of the sub-THz radiation are due to the manufacture deterriations, as mentioned in Section \ref{3dpm}. As a result, we propose the usage of vector space error metrics to describe the magnitude of mispositioning, instead of conventional distance metrics. By doing so, the beam management process will differ for each transceiver and each 3DPM. On the other hand, the distance $y$-axis affects the performance directly proportional as expected. While the EVM for $\pm \delta$ offset on $x$ alignment shows nearly 4 dB deviation for 164 GHz and 182 GHz, the deviation dramatically rises up for $\pm 2 \delta$ offset along with phase error.

\begin{figure*}[t] 
\centering
    \includegraphics[width=1\linewidth]{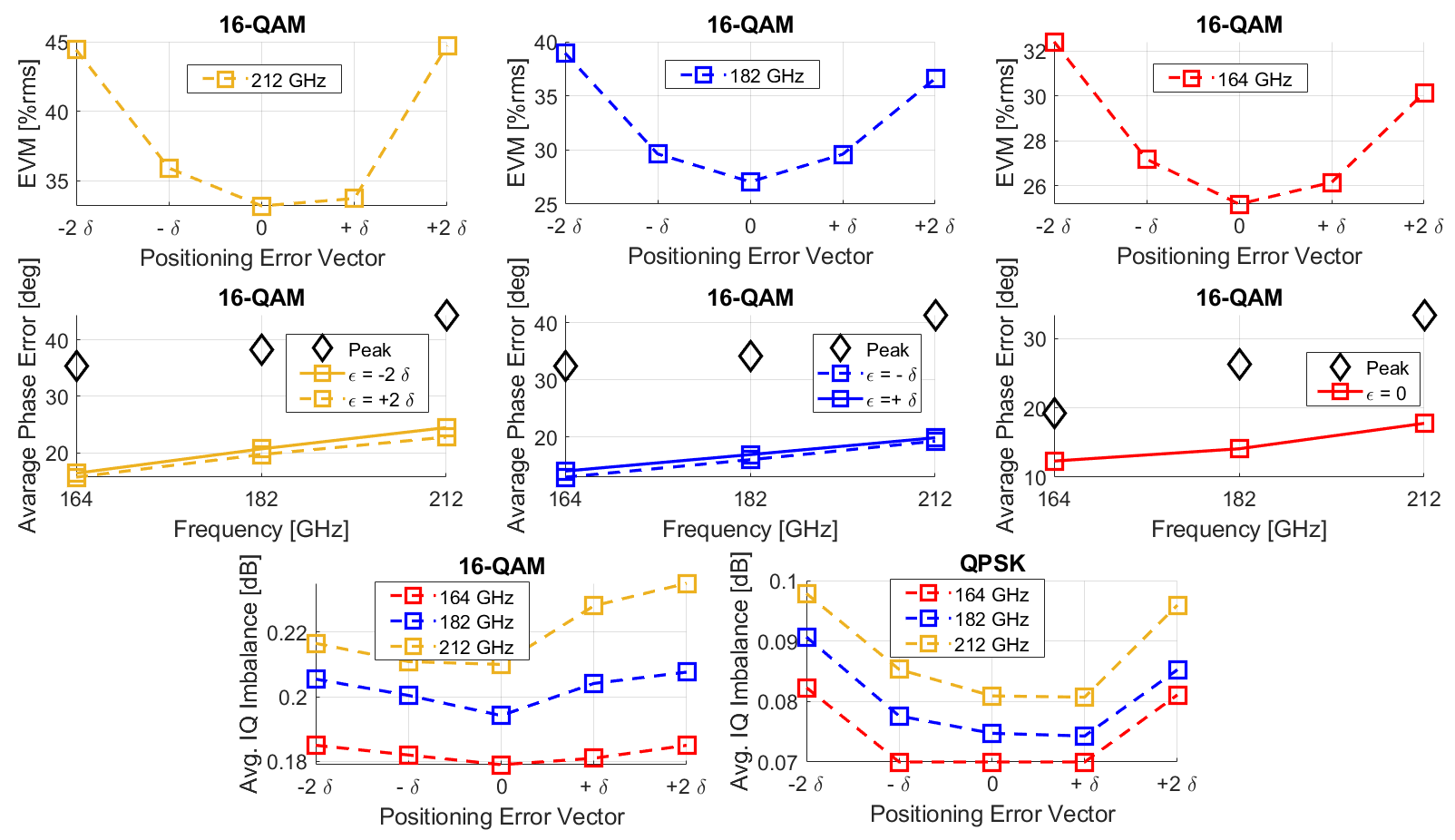}
    \caption{IQ-based measurement evaluation in EVM, Phase Error and IQ-imbalance for different $\delta$ positioning vectors that was conducted in 3 frequencies of the sub-THz region.}
    \label{res}
\end{figure*}
The IQ imbalance measurements for both QPSK and 16-QAM are shown in Figure \ref{res} which is taken in the approximately 19 dB SNR. Small distortions result in higher IQ imbalance in 16-QAM, this contradiction can be explained by the SNR estimation error on the 16-QAM constellation. Hereafter, the importance of the normalization reference of EVM shows itself, by taking the reference as the maximum magnitudes of the constellation makes the comparison more difficult. 

182 GHz is one of the peak attenuation, out of THz window frequency point where the water absorption, unlike the other two operating points. Yet, it is viable for both dry air and radiating near field applications. The test with mispositioning shows that the average phase error and peak distance show correlated behavior with 164 GHz and 212 GHz. Hence, it opens a window of use for the 182 GHz in some applications, \textcolor{black}{especially for the indoor environments as demonstrated in this study. Frequency selectivity throughout the experiments can be attributed to factors like voltage and phase noises on LO, unfiltered IF transmission components, and resonance efficiency variations across a wideband antenna.} 

A key point conclusion of the measurement is that the mispositioning impact is not caused by the path loss and molecular absorption as the $d_y$ is constant throughout the tests. In the case of large distance applications, wavefront manipulation with beam focusing is inevitable whereas in mobility, beam focusing and steering are indispensable. Note that in the sub-THz region, MIMO and RIS remain effective for mitigating high absorption loss, while beamforming offers limited assistance in contrast to mm-Wave and sub-6GHz. Nevertheless, state-of-the-art solutions involve beamfocusing and beam shaping closes this deficit. 

It \textcolor{black}{is} shown that phase and amplitude distribution in spatial geometry is not a deterministic behavior for sub-THz systems. This emphasizes that meter-level accuracy for sub-THz systems is not sufficient. The solution for high accuracy localization is possible for GNSS based methods such as real-time kinematic positioning (RTK), however, it is not an easy task for GNSS-denied areas as previously mentioned. MTL based localization methods will be fed from the CSI to not only demonstrate self-positioning but also strengthen the physical link.
\begin{table}[] 
\caption{Measurement Parameters}
\centering
\resizebox{0.35\textwidth}{!}{%
\begin{tabular}{l|l}
\hline
\textbf{Parameters}                         & \textbf{Value}           \\ \hline
Operating Frequencies {[}GHz{]}    & \{164, 182, 212\} \\
Number of Frames                   & 1500            \\
Conversion Loss {[}dB{]}           & 10 (dB)         \\
Modulation                        & QPSK, 16QAM     \\
$\delta$ {[}m{]}                   & 0.022 (m)       \\
$d_y$ {[}m{]}                      & 0.21 (m)        \\
Position Resolution                      & 0.001 (m)        \\
IF Sampling rate {[}Gbit s$^{-1}${]} & 8               \\
Beam Divergence {[}$^\circ${]}          & $\pm 7$         \\
$\theta_{\textrm{3dB},E}$ {[}$^\circ${]}         & $8.9$           \\
$\theta_{\textrm{3dB},H}$ {[}$^\circ${]}         & $10.28$         \\ \hline
\end{tabular}
}
\end{table}

\section{Open Issues}
While the revolutionizing aspect of 3DPM requires new definitions of channel state information (CSI) on THzCom fading, the single-slope path loss (SSPL) model is still reasonable to interpret the large scale fading for sub-THz channel modeling. \textcolor{black}{SSPL on THz can model the path loss for the short distance range in the simplest way \cite{tekbiyik2019statistical}. As the distance increases, atmospheric absorption-based models can represent path loss more accurately, despite an increase in complexity \cite{han2022molecular}. However, for indoor applications or outdoor scenarios where beam recovery is assisted by RIS, SSPL can be used to model received signal strength straightforwardly.} Furthermore, sub-THz generation and detection can be further explored in plasma wave devices.

Using sub-THz to prevent diffraction with Bessel beams is feasible. However, beam intensity is distance-dependent and constrained by phase velocity limits. Nevertheless, a finite number of rings of the beam can be utilized for tolerable distance in the sub-THz band. Once the phase velocity stabilized, however, it can be exploited with traveling waves of leaky wave antennas that guide its radiation along the length of the waveguide. This allows the 3D spectrum sensing (SS) localization with THzCom without requiring extra information \cite{ghasempour2020single}. 

\subsection{The Role of AI}
In addition to time and angle varying sensitivity due to the directive electromagnetic field of the sub-THz antennas, the dielectricity and roughness of the reflector surface disperse the wave. Hence, the division of outdoor, indoor and human body analyses is not well justified and misguides easily. AI based combinatorial optimization is a unique way to provide flexibility in fast response - high refresh rate - accuracy triangle in both estimation and detection.

The need for dimension reduction for 3D CSI can be done via principal component analysis (PCA) or data-driven methods. In radio localization, coordinate mapping and ranging optimization are other areas in which discriminative AI models can be utilized. The high rate data with nonlinear correlation in features can be exploited in latent space with an autoencoder for data compression. Getting out of the low GNSS sampling rate limitations with GNSS-denied localization methods, brings the THz sampling constraints forward in radio localization steps of 6G. Leveraging the spatial correlations to learn 3DPM is another opportunity to take advantage of AI \cite{kallehauge2023delivering} in the same manner as channel chart learning for radio mapping. 

\subsection{Even more Hardware Gaps}
THzCom is a research domain intricately tied to hardware, specifically pertaining to antennas and source generators. However, there are lingering aspects that have yet to be tackled in this field of study.
 
An untouched hardware related problem is that on-board local clock and LO. From the communication perspective, digital synchronization is possible, yet it is a major challenge in the sensing aspect. Even though GNSS serves a nano accuracy universal clock in a large coverage, it suffers from multiple difficulties that will affect the performance of THz applications. Other than the previously mentioned ART impact on front end chip, low refresh rate with 1 Hz and inaccessibility for out-of-coverage regions are critical constraints and to date, no investigations have shed light on this subject for ISAC.

Regarding the source generators, phase noise is still a bottleneck for CMOS based approaches in THz generators. Even though current extenders can reach -90 dBc/Hz in the sub-THz band with injection locking, it is either far from the ultra-wideband or high energy consumption of the voltage control oscillator.

\subsection{Integrated Sensing and Communication}

Merging the active and passive sensing applications with the communication link is a key enabler for spectral and energy efficiency in the next generation networks. Orthogonal frequency-division multiplexing (OFDM) offers robustness against MPC in wideband radar and communication, presenting an opportunity for active sensing. However, it is more susceptible to interference compared to frequency modulated continuous wave (FMCW). While FMCW is less robust against MPC than OFDM, hybrid waveform techniques and orthogonal time frequency space (OTFS) for \textcolor{black}{integrated localization and sensing} in sub-THz frequencies are emerging as promising approaches to leverage the spread characteristic of sub-THz.

\textcolor{black}{Our main vision for THz in ISAC is the following: THzCom inherently provides positioning information, and the position itself is a THz signal quality indicator. This radar-like character is a key potential of the THz systems. Even though the sensors for real-time environment monitoring in communication are the foundation of ISAC, the evolution of ISAC involves shifting the focus from sensing capabilities to a communication perspective.}

\textcolor{black}{Our study shows that THz localization for ISAC brings many opportunities. For example, it can enhance security and privacy for service users by utilizing positional data. Low-threat users can free up valuable resources, while high-threat users have the option to trade off signal quality indicators discussed in our demonstration.}

\section{Conclusion}

In conclusion, the unique properties that ISAC can bring into 6G and the availability of sub-THz bands for this technology in next generation communication \textcolor{black}{are} investigated and the opportunities in radio localization are listed. The significance of THz positioning in localization \textcolor{black}{is} discussed and a real-time sub-THz experiment \textcolor{black}{is} conducted to ground these arguments. It \textcolor{black}{is} observed that mispositioning leads to an unfavorable impact resolutely for all bands operated, 164 GHz, 182 GHz and 212 GHz, yet it reveals a wide range of subjects that will reinforce the localization process with an unusual twist. Lastly, future research challenges and opportunities have been discussed in this field. Overcoming the technical challenges and exploring further research avenues will pave the way for the widespread adoption of sub-THz based localization and sensing systems.

\bibliographystyle{IEEEtran}
\bibliography{bibliog}
\newpage
\section*{Biographies}
\begin{IEEEbiography}{ERAY GÜVEN}
received his B.S. degree in Electronics and Communication Engineering at Istanbul Technical University, Turkey in 2021. He is currently studying for a Ph.D. degree in electrical engineering at Polytechnique Montr\'eal, Montr\'eal, Canada.
\end{IEEEbiography}

\begin{IEEEbiography}{GÜNEŞ KARABULUT KURT}
is currently an Associate Professor of Electrical Engineering at Polytechnique Montr\'eal, Montr\'eal, Canada. She is a Marie Curie Fellow and has received the Turkish Academy of Sciences Outstanding Young Scientist (TUBA-GEBIP) Award in 2019. She received her Ph.D. degree in electrical engineering from the University of Ottawa, ON, Canada. She is a member of the IEEE WCNC Steering Board. 
\end{IEEEbiography}

\end{document}